\documentclass[twocolumn,aps,prd,epsf]{revtex4}
\usepackage[dvips]{graphicx}
\usepackage{epsfig}
\usepackage{color}
\usepackage{amsmath}

\newcommand{\be}{\begin{equation}}
\newcommand{\ee}{\end{equation}}
\newcommand{\bea}{\begin{eqnarray}}
\newcommand{\eea}{\end{eqnarray}}

\begin{document}

\title{The QCD string tension curve, the ferromagnetic magnetization, \\ and the quark-antiquark confining potential at finite Temperature}
\author{P. Bicudo}
\email{bicudo@ist.utl.pt}
\affiliation{CFTP, Departamento de F\'{\i}sica, Instituto Superior T\'ecnico,
Av. Rovisco Pais, 1049-001 Lisboa, Portugal}
\begin{abstract}
We study the string tension as a function of temperature, fitting the SU(3) lattice QCD finite temperature free energy potentials computed by the Bielefeld group. We compare the string tension points with order parameter curves of ferromagnets, superconductors or string models, all related to confinement. We also compare the SU(3) string tension with the one of SU(2) Lattice QCD. With the curve providing the best fit to the finite temperature string tensions, the spontaneous magnetization curve, we then show how to include finite temperature, in the state of the art confining and chiral invariant quark models. 
\end{abstract}
\maketitle

\section{Introduction}

Motivated by the discovery of the Quark-Gluon Plasma at the CMS at CERN, and RHIC at BNL laboratories and by the future experiments at RHIC-II, LHC and FAIR,  quark model computations are presently addressing with more and more detail chiral symmetry breaking and deconfinement at finite temperature or finite density. Here we show how to upgrade the confining quark-antiquark potential, with a string tension $\sigma $ function of the temperature $T$.

In the 80's a new quark model, the Chiral Quark Model ($\chi$QM), was developed
\cite{Yaouanc,Adler:1984ri,Bicudo_thesis}, 
including not only the quark confining potential of the early quark models, but also the spontaneous breaking of chiral symmetry of the Nambu and Jona-Lasinio model. 
Although the $\chi$QM can only be approximately derived from QCD in the Coulomb gauge hamiltonian formalism, although it has not yet been fully calibrated to the finest details of the hadronic spectrum, and although Lorentz invariance is only present in the kinetic energy, it is nevertheless the only QCD inspired model able to explicitly include both quark confinement and quark-antiquark vacuum condensation.  Thus the  $\chi$QM is the most comprehensive QCD model, adequate to study any possible system of quarks, gluons, or hadrons, and to provide, at least, a qualitative answer to the questions one may ask to QCD.

So far the $\chi$QM studies of chiral symmetry breaking at finite temperature
or density have assumed a temperature independent string tension, {\em i. e.}  the same confining potential at all temperatures, since the work of
Yaouanc, Oliver, P\`ene , Raynal, Jarfi and Lazrak
\cite{Yaouanc2}. 
Within this approximation, many studies 
of chiral symmetry have found that chiral symmetry breaking is maintained
at all temperatures. The first calculations at finite temperature or at finite density are already 20 years old
\cite{Yaouanc2,Bicudo:1993yh}, 
but presently finite temperature or finite density are interesting
many authors, including the S\~ao Paulo group  
\cite{Battistel:2003gn,Antunes:2005hp},
the Graz group 
\cite{Glozman:2007tv},
the Pittsburgh group
\cite{Lo:2009ud} 
and others. 
Only Lo and Swanson using ring diagrams, {\em i. e.} 
quark loops beyond the ladder approximation, 
has been able to find chiral symmetry restoration
at a finite critical temperature $T_c$,
since the quark loops effectively affect the confining potential.

\begin{figure}[t!]
\resizebox{0.5\textwidth}{!}{%
\includegraphics{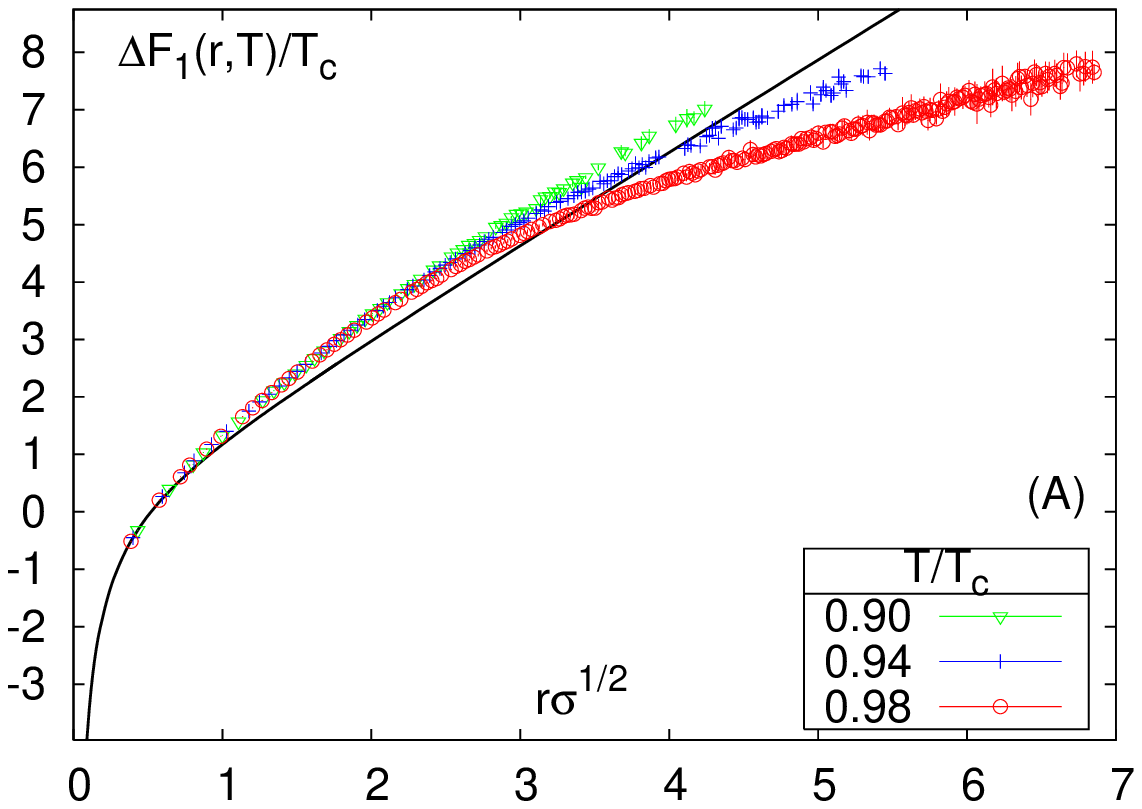}
}
\resizebox{0.5\textwidth}{!}{%
\includegraphics{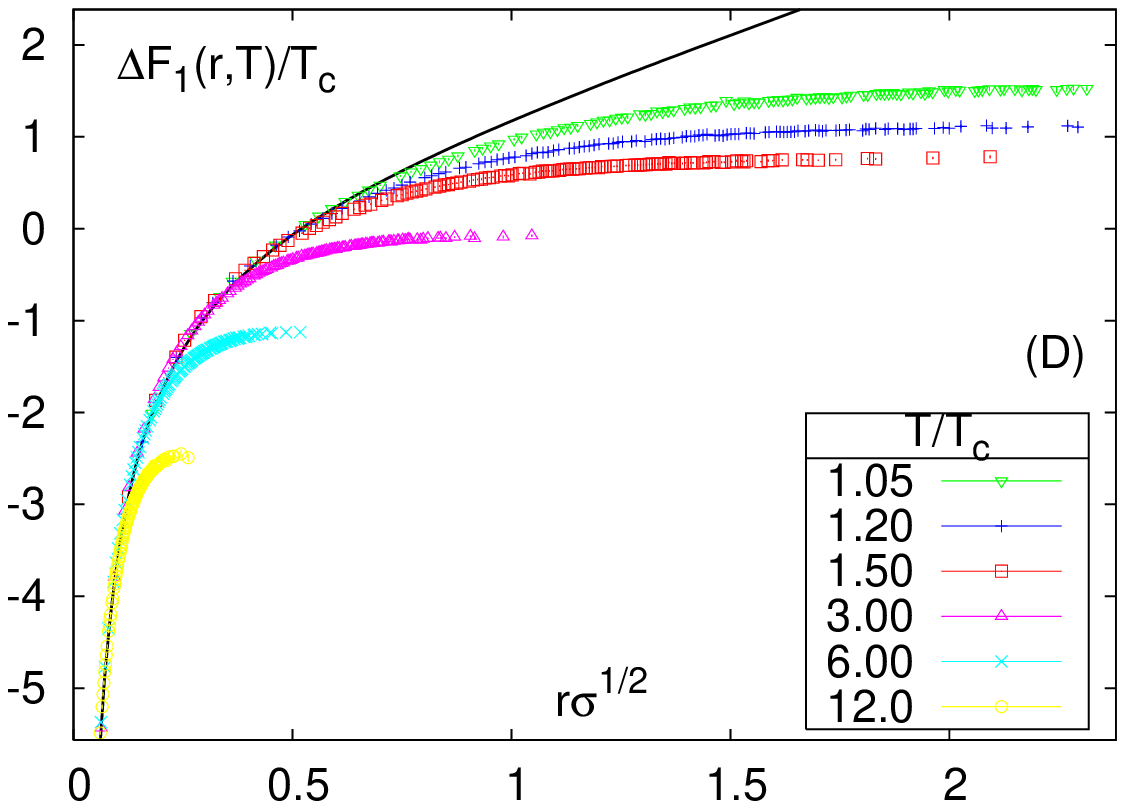}
}
\caption{\label{F1Kacz}
We show examples finite temperature static quark-antiquark potentials, in particular the  $T<T_c$  and $T>T_c$ 
Lattice QCD data for the free energy $F_1$, thanks to 
\cite{Doring:2007uh,Hubner:2007qh,Kaczmarek:2005ui,Kaczmarek:2005gi,Kaczmarek:2005zp}Olaf Kaczmarek et al.
 The solid line represents the $T=0$ static quark-antiquark potential. In this paper we discuss the use of the free energy as a finite temperature quark-antiquark potential.}
\end{figure}

Nevertheless the finite temperature static quark-antiquark free and
internal energies at some finite temperatures $T$ 
have already been computed in SU(3) lattice QCD, by the Bielefeld 
group
\cite{Doring:2007uh,Hubner:2007qh,Kaczmarek:2005ui,Kaczmarek:2005gi,Kaczmarek:2005zp}, 
and they are available to be included in  the $\chi$QM. 
In particular, the first part of the quark-antiquark potential to suffer
finite temperature effects is the long distance confining part.
In what concerns the short range Coulomb part of the quark-antiquark potential, it survives up to any temperature. Moreover the short-range
Coulomb potential is included in the ultraviolet renormalization program 
of QCD, thus it is relevant for the quark mass renormalization, but 
can be neglected for chiral symmetry breaking
\cite{Bicudo:2008kc}.
While the Coulomb potential at finite temperature is quite relevant
for the heavy quarkonia, since even slightly above the deconfinement
critical temperature $T_c$ the $J/\psi$ and $\eta_c$
remain bound
\cite{Bicudo:2008gs,Bicudo:2009pb}, 
for the binding of light quarks the confining part of the quark-antiquark
potential, parametrized by the linear string tension, is crucial.

\begin{figure}[t!]
\hspace{0cm}
\includegraphics[width=0.9\columnwidth]{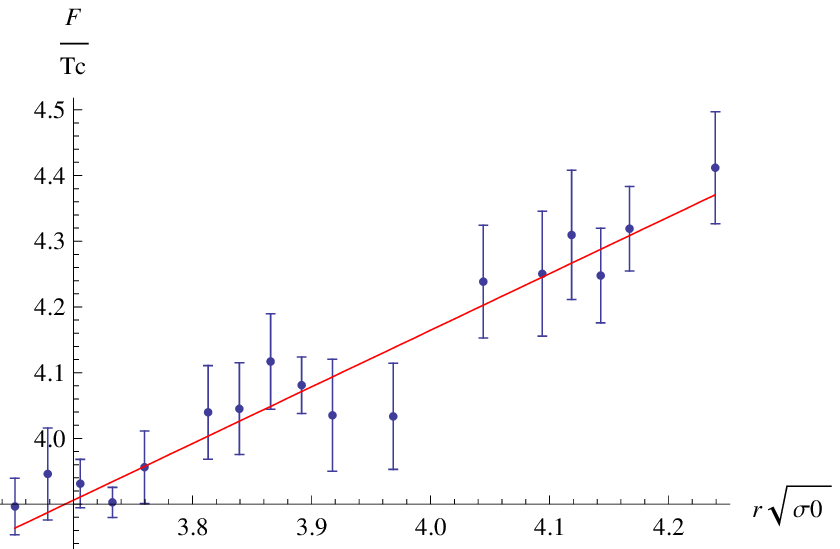}
\includegraphics[width=0.9\columnwidth]{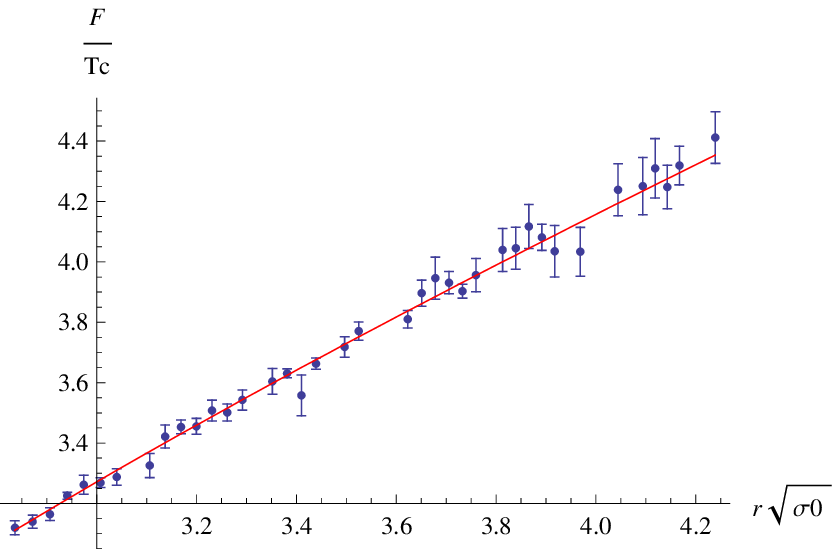}
\caption{Fits of the Bielefeld's group free energy at $T<T_c$, here in the case of $T =0.90 T_c$, in the top we use a linear fit and in the bottom we use a linear + coulomb fit.
We cut the low distance, or infrared,  part of the lattice data shown in Fig. \ref{F1Kacz}, in such a way that the fit is independent as possible on the cutoff. Similar cuts and fits are applied for the other temperatures. 
\label{fitlongdistance}}
\end{figure}

Moreover, although 
the Polyakov loop is presently the preferred order parameter for
deconfinement
\cite{McLerran:1980pk,McLerran:1981pb}, 
other order parameters may be used,
\cite{Zantow:2001yi,Zantow:2001yf}
and in particular the string tension is a possible order 
parameter for the deconfinement phase transition.

Thus here we specialize in the finite temperature string tension
\cite{Kaczmarek:1999mm}, 
and show how to include the results of finite $T$ Lattice QCD
in the $\chi$QM,  including finite $T$ in the string tension of the
confining part of the quark-antiquark potential. 
This works continues the study of the SU(3) string tension,
computed for the first time by the Bielefeld Group, 
\cite{Kaczmarek:1999mm}.

\begin{table}[t!]
\begin{tabular}{c|ccc}
\hline
$T / T_c $ & $\sigma$  & IR cutoff & $ \chi^2 /$ dof\\
\hline
0.00 & 1.545 $\pm$  0.002 & 1.0 &  - \\
0.90 & 0.861$\pm$ 0.072 & 3.6 & 0.52\\ 
0.94 & 0.587 $\pm$ 0.028 & 4.3 & 1.96 \\
0.98 & 0.429 $\pm$ 0.005 & 4.1 & 0.22\\
1.00 & 0.00 $\pm$  0.000 & - & - \\
\hline
\end{tabular}
\caption{ Fits of the string tension in units of $ {T_c}^{-1} {\sigma(0)}^{-1/2}$ for the
longer distance part of the Kackzmarek {\em et al} lattice data,
using the ansatz $ F(r)= c + \sigma r $.
\label{table1}}
\end{table}

In Section II we fit the SU(3) string tension from the long distance part of the 
recent static quark-antiquark free energy from the lattice QCD results of the 
Bielefeld Group, and we compare the SU(3) string tension with the SU(2) string tension.  In section III we compare the string tensions with the order parameter curves of ferromagnets, superconductors or string models, all related to confinement, and find that the curve closer to the 
SU(3) string tension is the magnetization curve. In Section IV we use the 
curve inspired in the magnetization of ferromagnets,  to also
compute the entropy string tension, the internal energy string tension, and
to estimate the finite $T$ string tension of the quark-antiquark potential.
In Section V, we conclude.

\section{Fitting the SU(3) and SU(2) Lattice QCD finite  $T$ string tensions with different ansatze}

This works continues the study of the SU(3) string tension,
computed for the first time by the Bielefeld Group, 
\cite{Kaczmarek:1999mm}, 
in a seminal paper also discussing in detail the finite temperature phenomenology. In this 2000 paper, the Bielefeld group computes
the string tension below $T_c$ utilizing colour averaged (over the
colour singlet and colour octet) Polyakov loops,
to extract colour averaged free energies. The results then obtained
confirmed a 1st order phase transition for the deconfinement
phase transition, as expected for SU(3) since the string tension
was found to remain finite at $T < T_c$, while it is vanishing
at $T>T_c$, with a small discontinuity at $T=T_c$.

More recently, the Bielefeld group computed
free energies for the colour singlet quark-antiquark static
system, utilizing gauge fixing. Here we fit the string tension
of these free energies, 
\cite{Doring:2007uh,Hubner:2007qh,Kaczmarek:2005ui,Kaczmarek:2005gi,Kaczmarek:2005zp}
recently computed by
Doring, Hubner, Kaczmarek, Karsch, Vogt and Zantow
in SU(3) Lattice QCD by , for $T<T_c$, see Fig. \ref{F1Kacz}.
For $T >T_c$ the fit is trivial since the confinement potential,
and the string tension, vanish.
To extract the string tension as  a function of the temperature $\sigma (T)$  we try different ansatze to fit the free energy, and
we cut the low distance part with an infrared cutoff. We choose the lowest possible cutoff, in the region where our fit is stable for changes of the cutoff. We also check that our fit is not too far from
the of $\chi/dof \simeq 1$. 
As different ansatze to fit the long distance part of the potential, we use
a constant and a linear term,
\be
F(r)= c(T)+ \sigma(T) r
\label{conslin}
\ee
or a constant, a coulomb and a linear terms
\be
F(r)= c(T) - {\alpha(T) \over r} + \sigma(T) r
\label{conslincoul}
\ee
or, as in the SU(3) Bielefeld fit \cite{Kaczmarek:1999mm}
and in the similar SU(2) fit of reference \cite{Digal:2003jc},
a constant, a linear and a Logarithmic terms
\be
F(r)= c(T) - a(T) \log(r) + \sigma(T) r
\label{conslinlog}
\ee
and in all of them we check the stability of the string tension.
We apply these ansatze to the free energies
at the different temperatures, and this is illustrated if 
Fig. \ref{fitlongdistance} for the case of $T=0.90 T_c$.
It occurs that the fit with a Logarithmic term of 
Eq. (\ref{conslinlog}) is not sufficiently stable, in the sense that 
whenever we increase the infrared cutoff,
the string tension $ \sigma$ change, and thus we abandon 
this ansatz. The other two fits of Eqs. (\ref{conslin})
and (\ref{conslincoul}) are stable, in the sense that
a distance $r_{cut}$ can be used as an infrared cutoff,
and that if we increase this infrared cutoff distance, the 
string tension is essentially unchanged.

\begin{table}[t!]
\begin{center}
\begin{tabular}{c|cccc}
\hline
\ $T / T_c $ \ & $\sigma$  & $\alpha$ & \ IR cutoff \ & $ \chi^2 /$ dof\\
\hline
0.00 & \ 1.395 $\pm$  0.005 \ & 1 & 0.3 & - \\
0.90 & \ 0.667 $\pm$ 0.012 \ & 10 & 2.8 &  0.51 \\
0.94 & \ 0.515 $\pm$ 0.008 \ & 11 & 2.8 & 1.56 \\
0.98 & \ 0.287 $\pm$ 0.002 \ & 14 & 3.0 & 0.21\\
1.00 & \ 0.00 $\pm$  0.000 \ & - & - & - \\
\hline
\end{tabular}
\caption{ Fits of the string tension in units of $ {T_c}^{-1} {\sigma(0)}^{-1/2}$ for the
longer distance part of the Kackzmarek {\em et al} lattice data,
using the ansatz $ F(r)= c - \alpha { \pi \over  12 \, r} +\sigma r $.
\label{table2}}
\end{center}
\end{table}

\begin{figure}[t!]
\hspace{0cm}
\includegraphics[width=0.9\columnwidth]{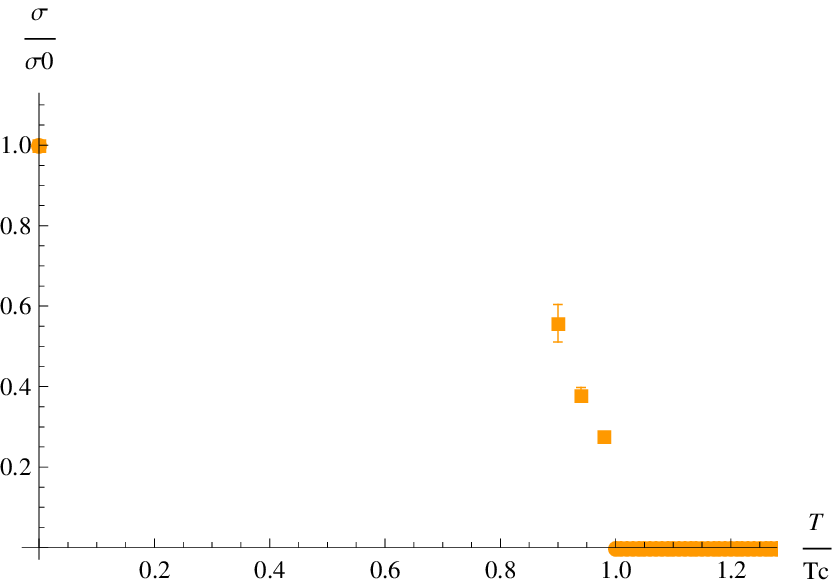}
\includegraphics[width=0.9\columnwidth]{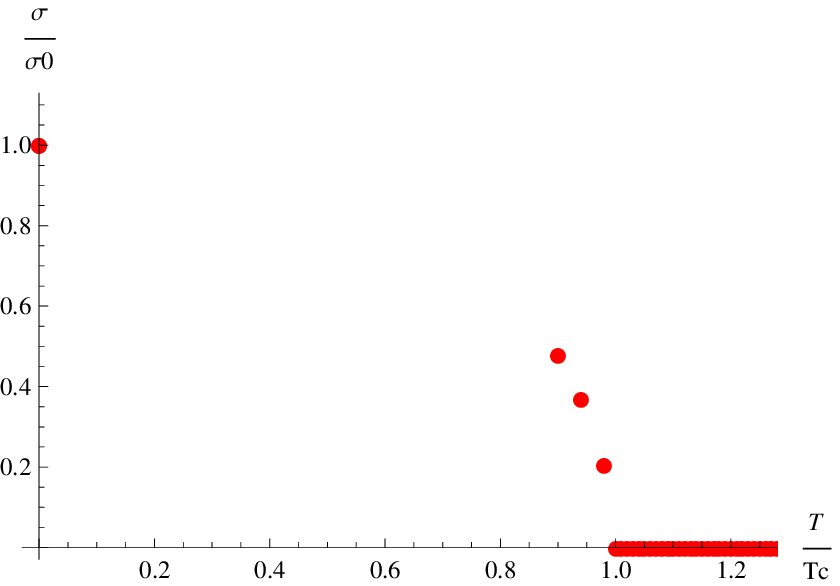}
\caption{The results of the fits, present in Table \ref{table1} (top)
and in Table \ref{table2} (bottom) , can be summarized in a dimensionless curve of $\sigma(T) / \sigma(0) $ as a function of $T / T_c$. The bottom fit,
of a constant plus a Coulomb plus a linear term has a greater stability
that the simpler top fit, and smaller error bars.
\label{fitalltemps}}
\end{figure}

The details on the fits are show in Tables \ref{table1}
and \ref{table2}. The simpler
constant plus linear fit of of Eq. (\ref{conslin}) needs
a higher cutoff $r_{cut}$  than the constant plus linear 
plus Coulomb fit of eq.  (\ref{conslinlog}).  The two different sets
of string tensions are depicted in Fig. \ref{fitalltemps}.
Notice that to the finite temperatures of  $T=0.90 T_c$,
$T=0.94 T_c$, and $T=0.98 T_c$ we add the cases of
$T=0 $ with the string tension $\sigma(0)$
and of $T > T_c$  with a vanishing string tension $\sigma(T)=0$.

It occurs
however that the string tensions computed with the
two different ansatze of Eqs. (\ref{conslin})
and (\ref{conslincoul}) differ, even when they are normalized
to $\sigma(T) \over \sigma(0)$. Notice that
the $T=0 $ potential has the Luscher term  $\pi \over  12 \, r$
and the same term is expected to also exist at finite
temperature,  the ansatz including a Coulomb term
of Eq. (\ref{conslincoul})  has a broader stability for changes of
the infrared cutoff $r_{cut}$, and produces
points with a smoother alignment. Thus of all our fits
of the string tension of the Bielefeld free energies,
the ansatz of Eq. (\ref{conslincoul}), with the fits of
Table \ref{table2} and Fig. \ref{fitalltemps} (bottom) 
appears to be the best.

\begin{figure}[t!]
\hspace{0cm}
\includegraphics[width=0.9\columnwidth]{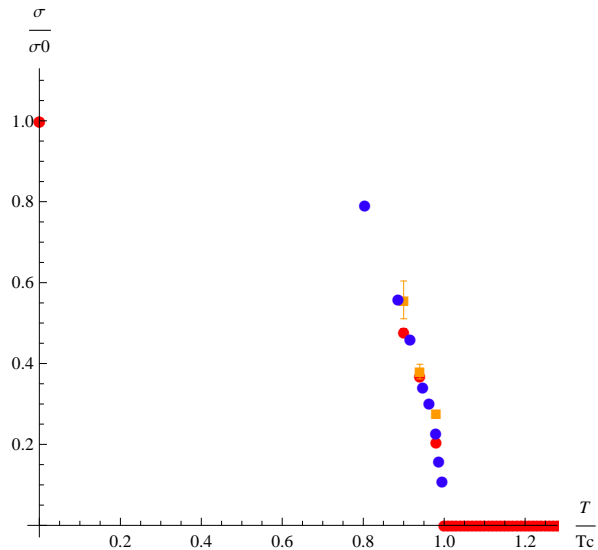}
\caption{The results of the fits, present in Table \ref{table1} 
and in Table \ref{table2}, together with the SU(3) fits in the year
2000 of the Bielefeld Group 
\cite{Kaczmarek:1999mm},  
summarized in a dimensionless curve of $\sigma(T) / \sigma(0) $ as a function of $T / T_c$. 
\label{AllSU3stringtensions}}
\end{figure}

In Fig. \ref{AllSU3stringtensions} we plot our fits, present in 
Table \ref{table1} and in Table \ref{table2},
together with the string tensions $\sigma$ that were fitted by the Bielefeld 
Group in 2000
\cite{Kaczmarek:1999mm}. They are all consistent, with our linear 
fit slightly larger than the Bielefeld fits in the year 2000,and with
our linear plus Coulomb fit only slightly lower than the Bielefeld fit. 
In Fig. \ref{andSU2}
we also compare the SU(3) string tension critical curve (bullets)
with the SU(2) string tension critical curve (squares) 
of reference \cite{Digal:2003jc}.
The SU(2) sting tension is smaller than the SU(3) one, and it corresponds 
to a larger critical exponent than the critical exponent of SU(3).

\section{Ferromagnetic, superconductor and string inspired ansatze for the QCD string tension curve}

The string tension $\sigma$ is crucial for the spontaneous breaking of
chiral symmetry, and for the light hadron spectra. 
Thus it is important to know the string tension $\sigma(T)$ at all temperatures, staring from $T=0$ and not just at temperatures 
of the order of $T_c$. 
Since in Section II the string tension was only computed
for a few temperatures, close to $T_c$, we now compare the string tension 
to similar other curves or order parameters, with the aim to propose an ansatz for $\sigma(T)$.

We remark that we expect for QCD a 1st 
order phase transition with a discontinuity at $T= T_c$, 
but with only a weak discontinuity. Thus we may use
as ansatze for the string tension curve, in the case
where we are interested in all the interval $ T \in [0, T_c]$
and not just in a narrow neighbourhood of $T_c$,
second order parameter curves. Indeed 
the points in Fig. \ref{fitalltemps} are close to second order, 
or continuous phase transition, with a critical exponent in the order 
parameter. Notice that we do not claim here that the transition is not the
expected 1st order one, we are simply concerned with the fit of the potential
to be used in the quark model. 

To find an ansatz for the string tension curve we first study
order parameter curves of physical systems related to 
confinement. The Ising model not only is a model of
ferromagnetism, but it is also a model of confinement,
and in particular the critical exponent of the SU(2)
string tension is similar to the three-dimensional
Ising model exponent. Another model of confinement
is the dual superconductor model, since the colour
electric flux tubes existing between confined quarks and
antiquarks are confined as vortices in superconductors. 
And colour electric flux tube confinement can be approximated,
in the limit of thin flux tubes, by string models. Thus we
may inspire ourselves in ferromagnetic materials, in the Ising
model, in superconductors either in the BCS model or in the
Ginzburg-Landau model, or in string models, to suggest 
ansatze for the string tension curve.

\begin{figure}[t!]
\hspace{0cm}
\includegraphics[width=0.9\columnwidth]{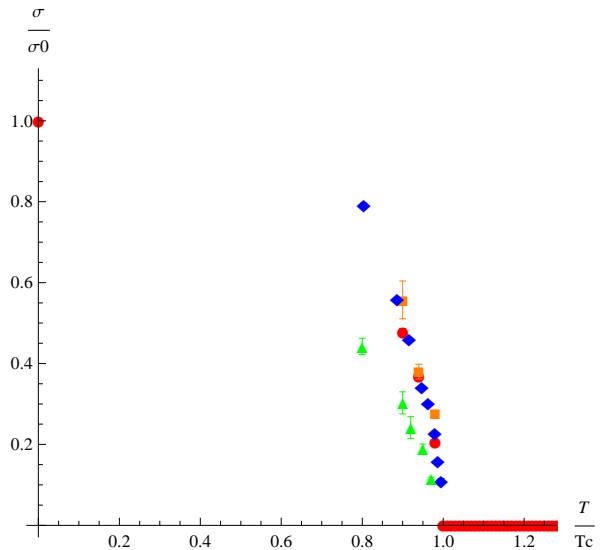}
\caption{Comparing the SU(3) string tension critical curve (bullets, squares and diamonds)
of Fig. \ref{AllSU3stringtensions}, with the SU(2) string tension critical curve (triangles) 
of Ref. \cite{Digal:2003jc}.
The critical curve of SU(2) is flatter, and it corresponds to a larger 
critical exponent 0.63 than the critical exponent $\simeq$ 0.5 of SU(3).
\label{andSU2}}
\end{figure}

\begin{figure}[t!]
\hspace{0cm}
\includegraphics[width=0.9\columnwidth]{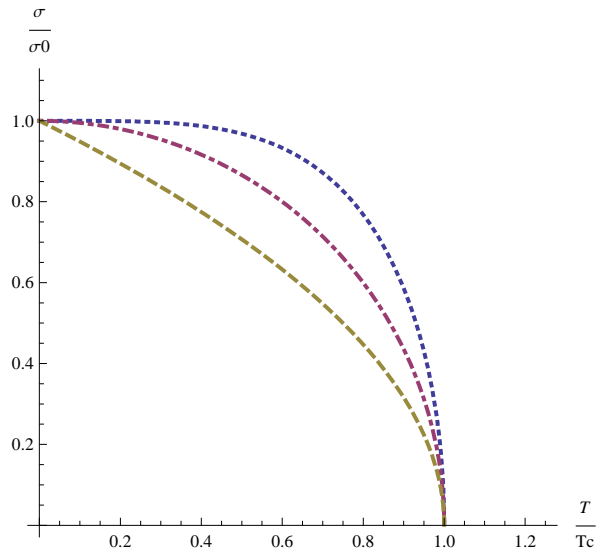}
\caption{We show the string tension curves 
$\sigma \over sigma(0)$ as a function of $T \over T_c$ 
of the parametric ansatze 
of Eq. (\ref{GL}) in a dashed line, Eq. (\ref{stringosc}) in a dot-dashed line , and Eq. (\ref{empirical})
in a dotted line. The solution of Eq. (\ref{massgap}) for a large Debye frequency 
$w_D$ is similar to the dashed line. 
\label{parametrics}}
\end{figure}

\begin{figure}[t!]
\hspace{0cm}
\includegraphics[width=0.9\columnwidth]{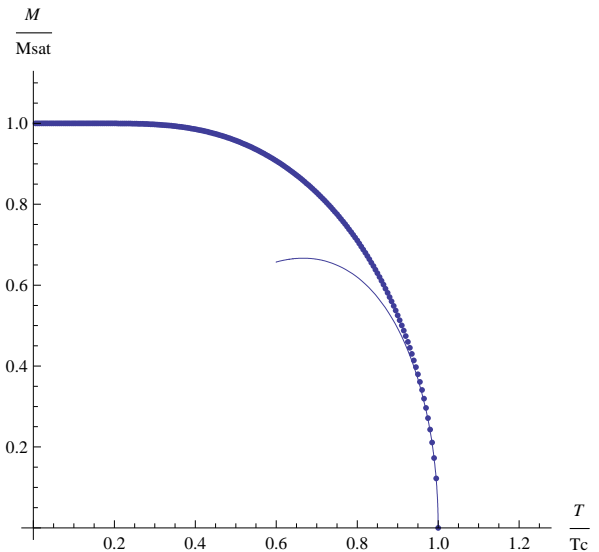}
\caption{The critical curve for $M \over M_{sat}$ as a function of $T \over T_c$ is show in a dotted solid line (one dot per each of our numerical fixed point solution). 
We find the critical curve solving Eq. (\ref{eqformagnetization})
with the fixed point method. The solution is essentially 
a flat constant curve for low $T$ but for $T\simeq T_c$ it behaves like the square root as in eq. (\ref{squarerrot}), also depicted here.
\label{magnetiz}}
\end{figure}

We order the different ansatze used in these related systems,
from the simpler to define, to the more sophisticated one.
The coherence length of a Ginsburg-Landau superconductor
\cite{deGennes}
is,
\be
{\xi \over \xi_0} = \sqrt{ 1 - {T \over T_c} }
\label{GL}
\ee 
the string tension of a finite $T$ string
\cite{Gao:1989kg}, 
extrapolated from $T \simeq T_c$ to $T \simeq 0$, is,
\be
{\sigma \over \sigma_0} = \sqrt{ 1 - \left(T \over T_c\right)^2}
\label{stringosc}
\ee 
the empirical fit to the penetration length
of a superconductor
\cite{FetterWalecka}
is,
\be
{\lambda \over \lambda_0} = \sqrt{ 1 - \left(T \over T_c\right)^4}
\label{empirical}
\ee 
the spontaneous magnetization of a ferromagnet 
\cite{FeynmanLS}
is the solution of the
algebraic equation,
\be
{M \over M_{sat}} = \tanh \left(  { T_c \over T}  {M \over M_{sat}}  \right) \ ,
\label{eqformagnetization}
\ee
and the mass gap of a BCS superconductor 
\cite{FetterWalecka}
is a solution of the integral equation,
\be
1 = g N(0) \int_0^{w_D} {d w \over w^2 + \Delta^2}
\tanh {\left(  w^2 + \Delta^2 \right)^{1/2}
\over 2 T  } \ .
\label{massgap}
\ee
We compare the curves of Eqs.
(\ref{GL}), (\ref{stringosc}), (\ref{empirical})
in Fig. \ref{parametrics}, and we show how to
plot the curve of Eq.  (\ref{eqformagnetization})
in Fig. \ref{magnetiz}. 
The mass gap solution of Eq. (\ref{massgap})
is not unique since it depends on two parameters,
the density of states $g N(0)$ and the Debye frequency $w_D$.
In the limit of a very small $w_D$, Eq. (\ref{massgap})
is equivalent to  Eq.  (\ref{eqformagnetization}). 
The opposite limit $w_D >> g N(0) $ occurs in real type I 
superconductors
\cite{FetterWalecka}, in which case the curve $\Delta(T)$, 
in dimensionless units, 
is quite similar to the circular curve of Eq. (\ref{stringosc}).
Thus we do not plot the solution of  Eq. (\ref{massgap})
in a separate curve.

\begin{figure}[t!]
\hspace{0cm}
\includegraphics[width=0.9\columnwidth]{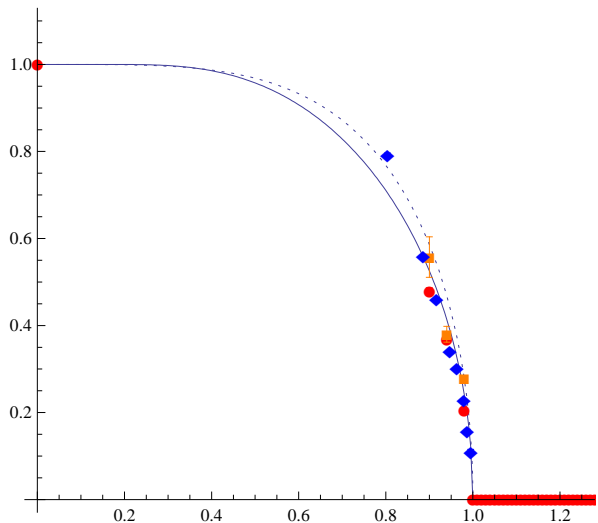}
\caption{Comparing the ferromagnet magnetization $M/M_{sat}$ critical curve (solid line) of Fig. \ref{magnetiz} with the SU(3) string tension 
$\sigma / \sigma(0)$  (bullets, squares and diamonds) of Fig. \ref{fitlongdistance}, both
as a function of $T\simeq T_c$. 
In this plot, $M/M_{sat}$  is quite close to $\sigma / \sigma(0)$. 
We also show (dotted line) the empirical curve of Eq. (\ref{empirical})
which provides the second
best fit of all the ones we tried.
\label{andsigma}}
\end{figure}

\begin{figure}[t!]
\hspace{0cm}
\includegraphics[width=0.9\columnwidth]{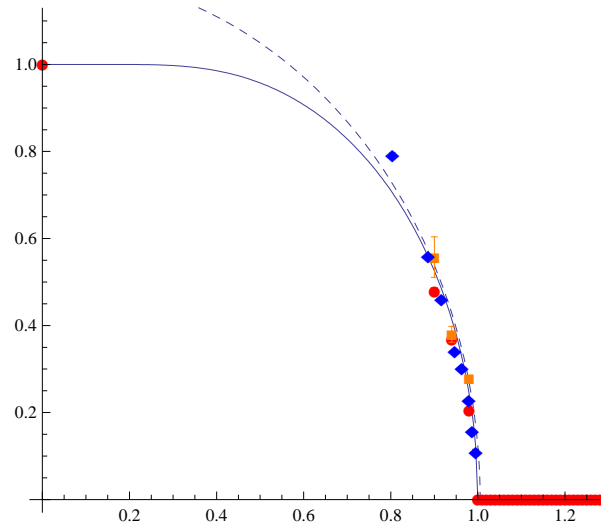}
\caption{Comparing the magnetization curve (solid line) with the SU(3) string tension critical curve (bullets) computed in Ref. \cite{Kaczmarek:1999mm}.
We also show (dashed line) the fit $1.21( 1- 0.990 (T/T_c)^2 )^{1/2}$ used in  Ref. \cite{Kaczmarek:1999mm} to measure the finite string tension at $T_c$ as an evidence for a 1st order phase transition.
\label{andSU31storder}}
\end{figure}

In Fig. \ref{andsigma} we plot the Magnetization, solution of Eq. (\ref{eqformagnetization}),
together with the string tensions $\sigma$ that fitted from 
SU(3) Lattice QCD data. The solution of Eq. (\ref{eqformagnetization})
provides our best fit of the string SU(3) tensions. We also show in
Fig. \ref{andsigma} our second best ansatz, the one of the empirical
penetration length of a ferromagnet in eq. (\ref{empirical}).  
In Fig. \ref{andSU31storder} we also compare the magnetization curve  
with the SU(3) string tension critical curve computed in Ref. 
\cite{Kaczmarek:1999mm}, and with
 the fit 
\be
{\sigma \over \sigma(0) }=1.21 \sqrt{ 1- 0.990 \left( T \over T_c \right)^2 }
\label{1storderfit}
\ee
 used in  Ref. \cite{Kaczmarek:1999mm} to measure the finite string tension at $T_c$ as an evidence for a 1st order phase transition. Both the fit
of Eq. (\ref{1storderfit}), similar to Eq. (\ref{stringosc})
but with a different norm and with a small temperature shift of $10^{-2} T_C$, showing
evidence for the 1st order phase transition, and the SU(3) string are quite close to our magnetization curve for $T$ close to $T_c$. But the magnetization
curve is the only one that extends up to the correct solution at $T=0$.
The magnetization curve is as close to our string tension curve as it
is close to the magnetization of a real ferromagnet.

Since the magnetization curve fits so well the SU(3) string tensions,
we now briefly review how it is derived.
The ferromagnetic spontaneous magnetization curve as a function of temperature is a text book curve, detailed for instance in the
{\em Feynman Lectures on Physics}
\cite{FeynmanLS}. 
Let us briefly review the simplest case, with
spins in the $1 \over 2$ 
representation. Then, in a magnetic field $B$, each spin has 
only two possible states with energies, $\pm \mu_0 \, B$ where $\mu_0 =
{g \over 2 } {e \hbar \over 2 m }$ is not the chemical potential
but it is the magnetic moment of a quantum spin. 
This leads to the magnetization,
\be
M = N \mu_0 \tanh { \mu_0 \, B  \over k \, T} \ .
\ee
The case that interests is the case of a ferromagnet, when there is no external field, and 
the magnetic mean field affecting the spin is proportional to the magnetization $M$.  $N \mu_0$ is the saturation magnetization $M_{sat} $. We then
get the algebraic Eq. (\ref{eqformagnetization}) equation for the magnetization curve,
\be
{M \over M_{sat}} = \tanh \left(  { T_c \over T}  {M \over M_{sat}}  \right) \ ,
\nonumber
\ee
the Eq. providing the best ansatz for our string tension curve. 

To analyze the curve solution to Eq. (\ref{eqformagnetization}),
we start by getting the extreme points of the curve solution to eq. (\ref{eqformagnetization}) , noticing that $\tanh \infty=1$, thus when $T=0$ we get ${M \over M_{sat}} =1$.
Also when the only solution of $x = \tanh x$ is $x=0$ thus when 
$T=T_c$ we have ${M \over M_{sat}} =0$ as expected. Moreover,
expanding the hyperbolic tangent close to $T \simeq T_c$, we get ,
\begin{eqnarray}
&& {M \over M_{sat}} \simeq  { T_c \over T}  {M \over M_{sat}} -  {1 \over 3} \left( { T_c \over T}  {M \over M_{sat}}\right)^3
\nonumber \\
& \Rightarrow & {M \over M_{sat}} \simeq   \sqrt{ 3{ T \over T_c} \left({ T\over T_c}  -1  \right) } \ .
\label{squarerrot}
\end{eqnarray}
This shows that in this case the critical exponent for the magnetization
$M$ is $1 \over 2$. This corresponds to a second phase transition since 
the magnetization is the first derivative of the free energy with regards to the chemical potential. For a complete solution of Eq. (\ref{eqformagnetization}),
we use the fixed point method, with $10^4$ iterations for each temperature $T$.
In Fig. \ref{magnetiz} we show the solution of Eq.  (\ref{eqformagnetization})
obtained with the fixed point expansion, and also the approximate solution
shown in of Eq. (\ref{squarerrot}) for $T \simeq T_c$.

It is striking that the curve $M( T/T_c) \over M_{sat}$, which is
derived from a simple two-state quantum spin system, not only
fits the magnetization curve of many ferromagnets, but also
fits essentially the string tension curve of finite temperature quenched SU(3)
QCD.
Possibly this occurs since confinement can also be partially understood
in the simplest two-state Ising Model. In particular Digal, Fortunato
and Petreczky
\cite{Digal:2003jc}
computed the string tension of quenched SU(2), also with fits of the
free energy, and found that the SU(2) string tension has the same
critical exponent 0.63 of the 3-dimensional Ising model. Thus in Fig. \ref{andSU2}
we also compare the SU(3) string tension critical curve (bullets)
with the SU(2) string tension critical curve (squares) 
of reference \cite{Digal:2003jc}.
The critical curve of SU(2) is lower than the SU(3) curve, 
and it corresponds to a slightly
larger critical exponent than the critical exponent $\simeq 0.5$ of SU(3),
obtained when neglecting the first order discontinuity at $T=T_c$. Nevertheless the
similarity between our SU(3) string tension and the the SU(2) and
Ising model string tensions suggests that the SU(3) confinement
is relatively simple. Indeed Lattice QCD is able to simulate 
comprehensively hadronic physics using a relatively small number 
of 100 to 1000 configurations, and this is only possible if SU(3)
confinement has relatively few degrees of freedom.
It may also be relevant for the understanding of SU(3) confinement,
that the mean field model of ferromagnetism provides a better fit 
to the SU(3) string tension curve than the Ising spin-spin interaction 
model. Also, when comparing with a the circular curve of 
Eq. (\ref{stringosc}), the SU(3) string tension curve is flatter at
the origin, and thus closer to the curve of Eq. (\ref{empirical}),
while the SU(2) string tension curve is steeper and thus closer
to the curve of Eq. (\ref{GL}).

Thus, based on our numerical results, and also confident in the
relative simplicity of SU(3) QCD confinement, and on the
small discontinuity of the string tension $\sigma(T)$ at $T_c$
we may use, in what concerns finite temperature quark models,
as an ansatz for the string tension,
\be
\sigma (T/T_c) \simeq \sigma(0) {M( T/T_c) \over M_{sat} } \ .
\label{anssigma}
\ee

\section{The free energy, the internal energy, and the finite temperature quark-antiquark confining potential}

\begin{figure}[t!]
\hspace{0.cm}
\includegraphics[trim = 22mm 0mm 0mm 0mm,width=1.09\columnwidth]{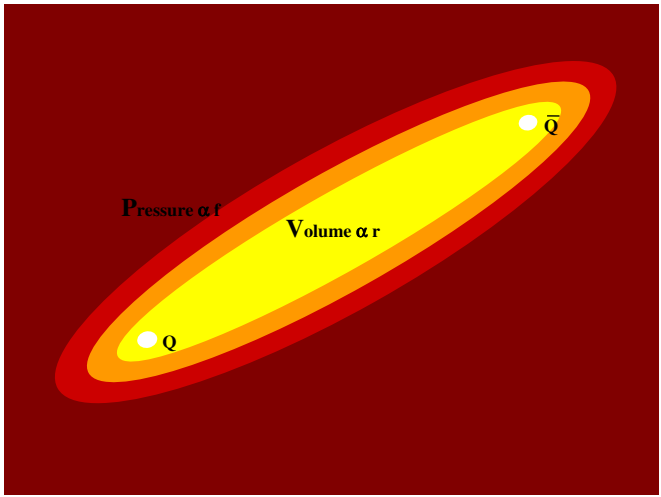}
\caption{A free illustration of a static flux tube, where we compare
the force $f$ to the pressure $P$, and the length $r$ to the volume $V$. 
\label{fluxtube}}
\end{figure}

In the $\chi$QM, the potential energy $V$ is used
in the mass gap equation to compute the quark constituent mass
and chiral symmetry breaking, and in the Salpeter equation
to compute the hadron spectrum.
However, in Lattice QCD, first the quark-antiquark free energy $F$ is 
computed from the pair of Polyakov loops $P$, 
\be
\mbox{Tr} \left\{ P(0)P^\dagger(r) \right\}= C \exp \left( - F \over kT \right) \ ,
\ee
and it is the free energy of the a static quark-antiquark pair
that we have studied in Sections II and III.
The potential energy of the quark-antiquark pair is related
to the quark-antiquark force,
\be
d V= - f \, dr 
\ee 
where a linear flux tube is assumed as in Fig. \ref{fluxtube},
and we can study the thermodynamics of the
flux tube using the equivalence $f \, dr = p \, dV$ for the
work of the quark-antiquark pair.
Then we get the usual thermodynamic relations, relevant at finite $T$,
for the free energy $F$,
\be
F (r)= - f \, d r  S \,d T \ , 
\ee
and for the internal energy $E$,
\be
E (r)= - f \, d r + T \, d S \ .
\ee
The quark potential $V$ is identical to the free energy $F$
in a reversible isothermal transformation with $dT=0$ , 
and is identical to the internal energy $E$ in a
reversible adiabatic transformation with $TdS=0$.

\begin{figure}[t]
\hspace{0cm}
\includegraphics[width=0.45\columnwidth]{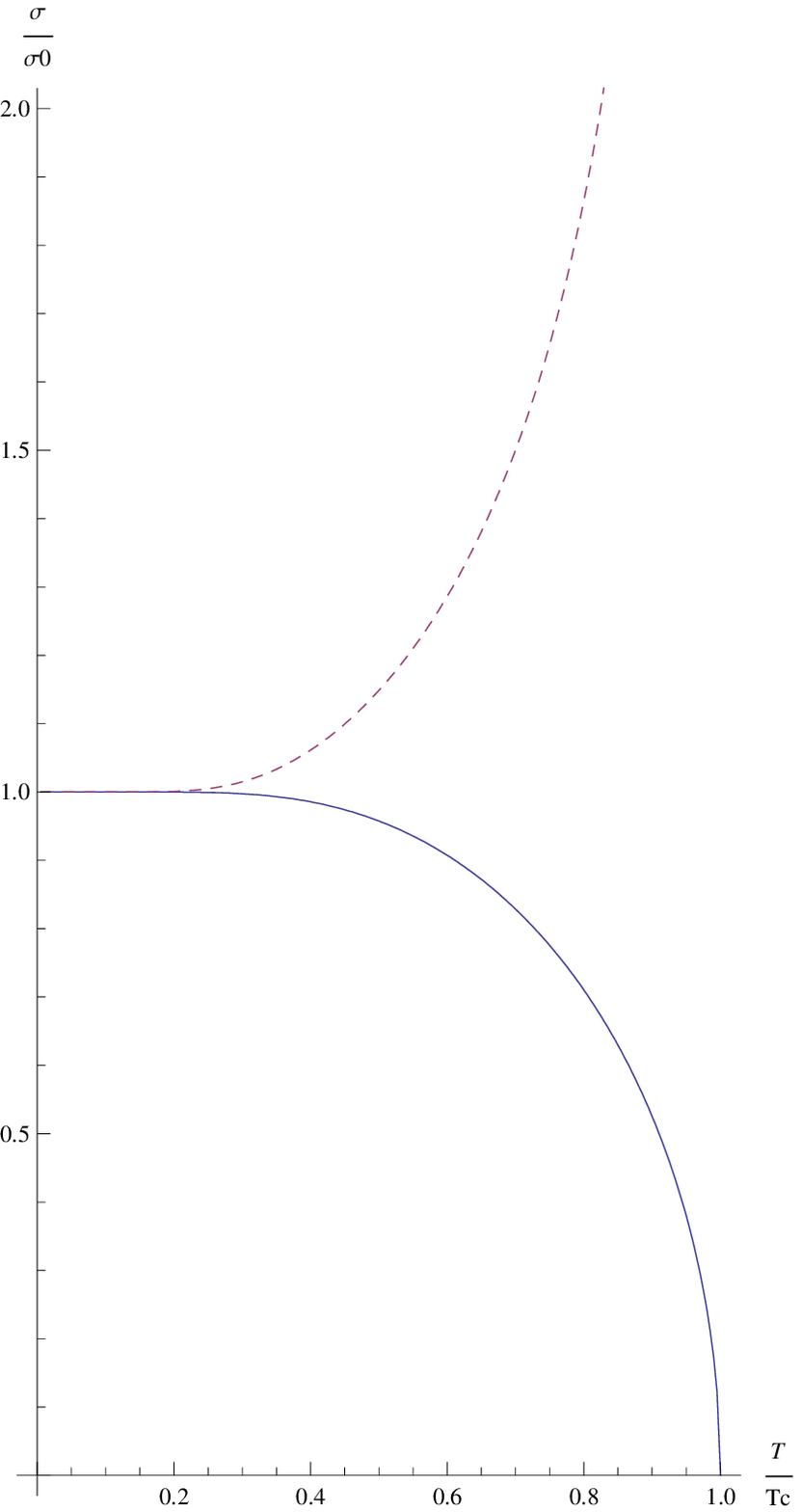}
\includegraphics[width=0.45\columnwidth]{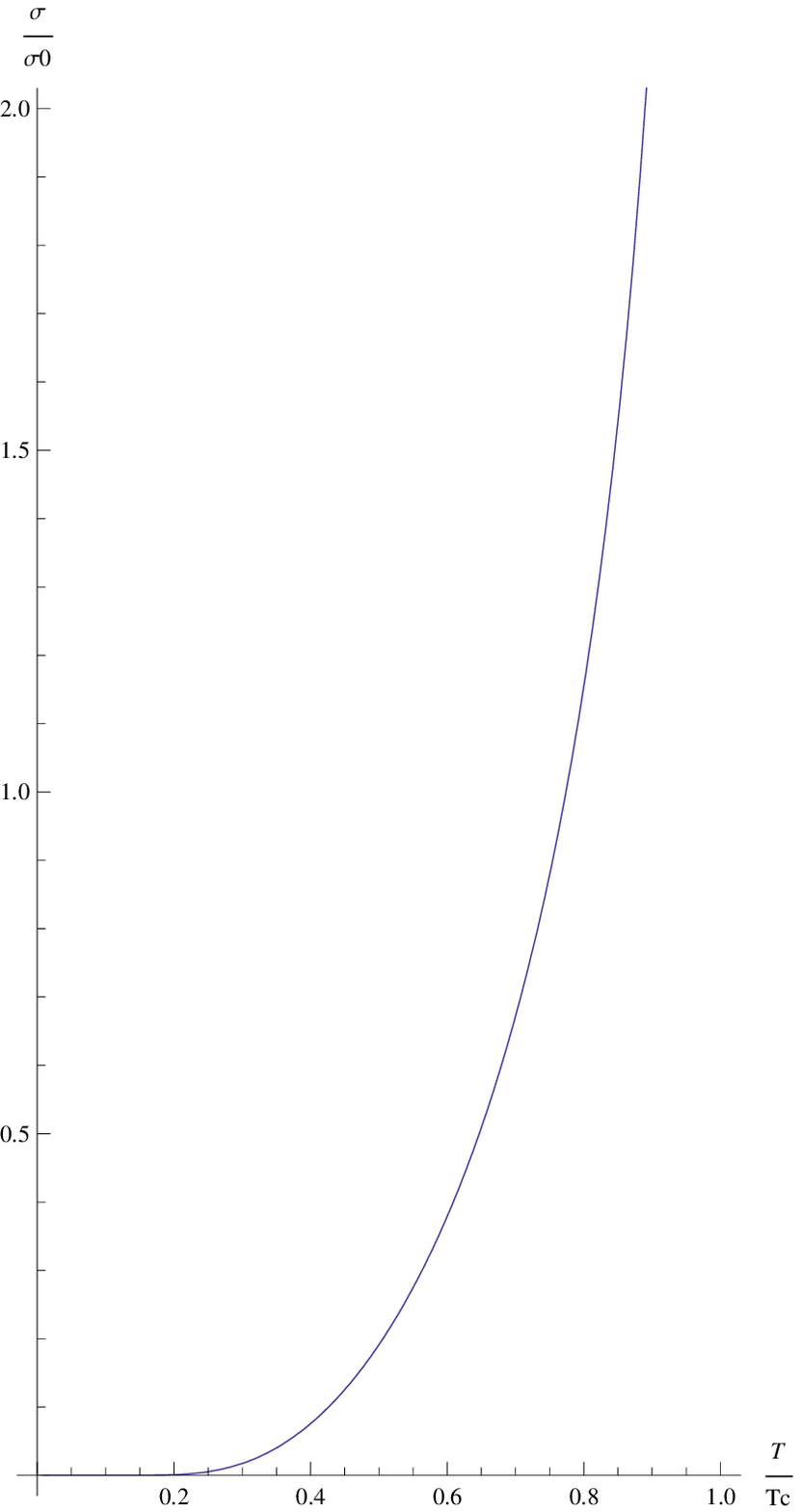}
\caption{Free energy (solid line) compared with the internal energy (dashed line) (left), entropy divergent at $T=T_c$ (right). 
\label{energies}}
\end{figure}

Since we have a good ansatz for the confining part of $F$, 
{\em i. e.} for the string tension, determined in Section III,
we may compute the string tension of the entropy $S$,
\be
S= - {\partial F \over \partial T} \ ,
\ee
and then also compute the string tension of the internal energy,
\be
E = F + TS \ .
\ee
In Fig. \ref{energies}, we compare the different string tensions.
Notice that the entropy diverges when $T \to T_c$, and this
is well known from the computations of the Bielefeld Group
at temperatures close to $T_c$. In particular the divergence is,
\be
\sigma_S \to {\sigma_0 \over 2 \, T_c}\sqrt{ 3 \over  {T \over T_c}-1   } \ ,
\ee
and thus the internal energy $E$ also diverges when $T \to T_c$,
this is clear in Fig. \ref{energies}.

\begin{figure}[t]
\hspace{0cm}
\includegraphics[width=0.45\columnwidth]{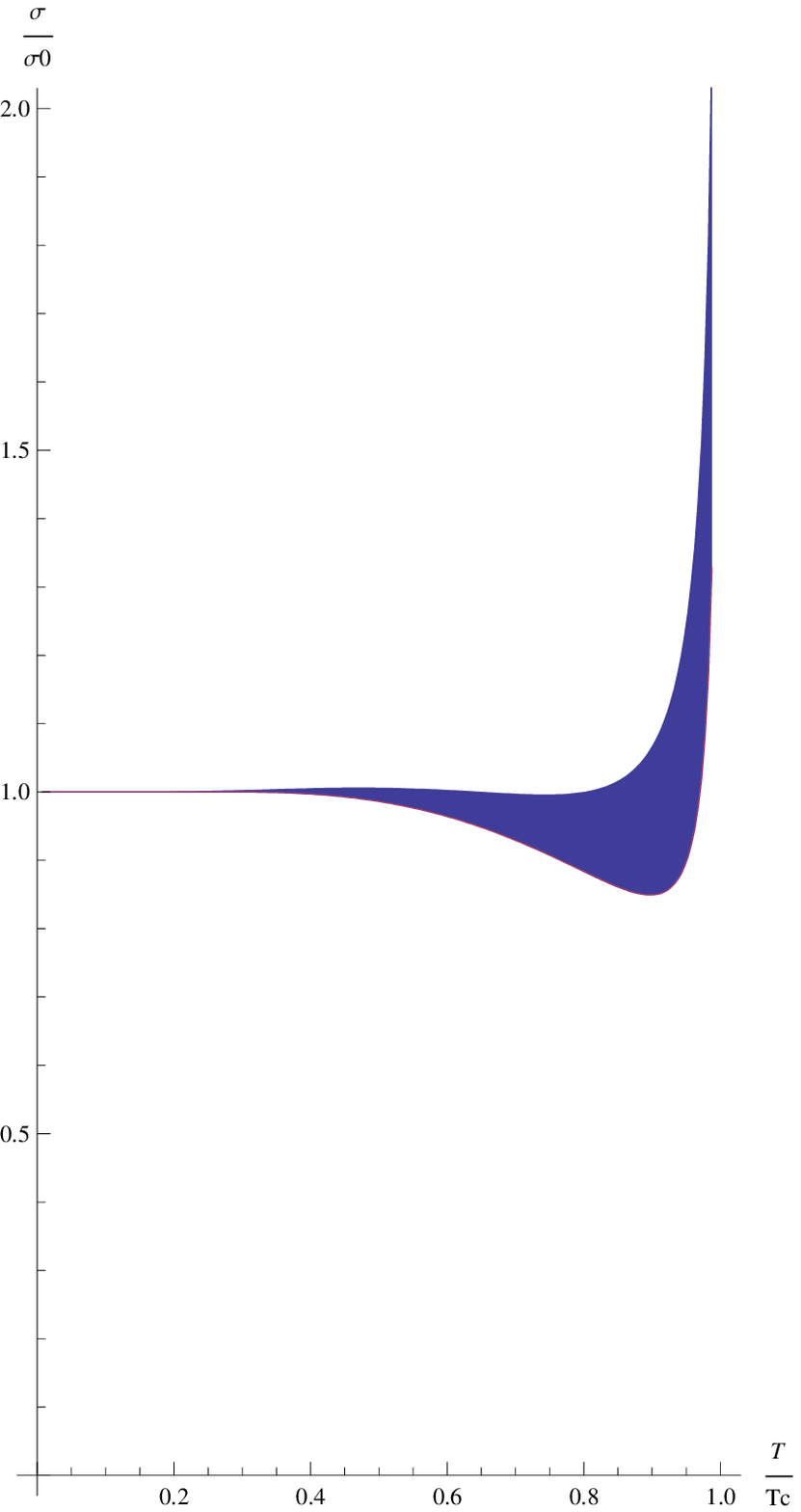}
\includegraphics[width=0.45\columnwidth]{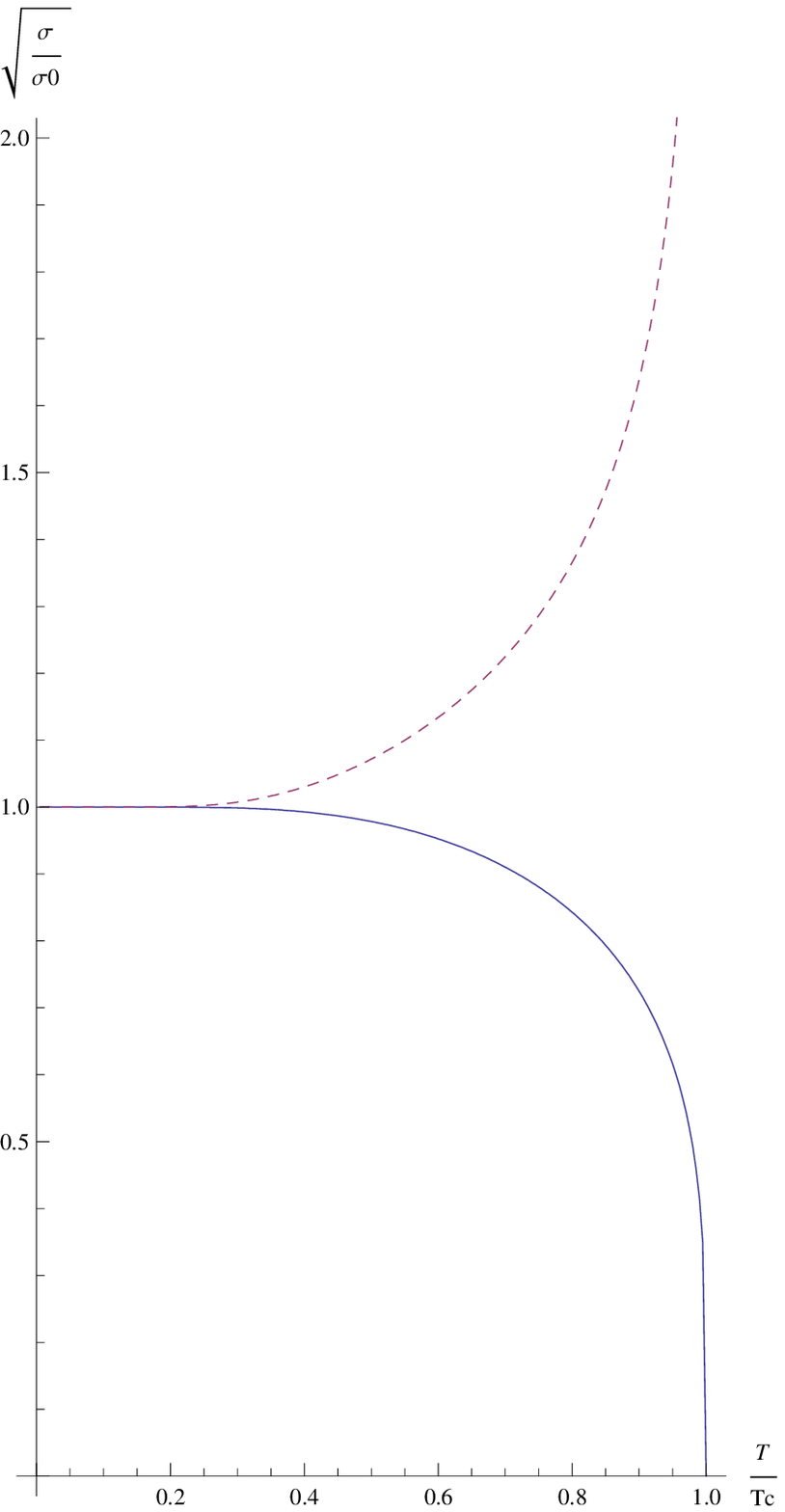}
\caption{Flattening the potential combining some internal energy to the free energy 
(left), flattening the free (solid line) and internal (dashed line) energies with a square root
(right). 
\label{flattening}}
\end{figure}

Now, in a hadron, or in the vacuum, it is not totally clear what
energy, either $F$ or $E$ to use as a the quark-antiquark potential. 
A main assumption of the $\chi$QM, and of any quark model, is 
that the static quark-antiquark computed in Lattice QCD can be also be used,
at least qualitatively, for light quarks. One assumes that the quark moves 
relatively slowly compared to the flux tube formation typical scale,
and that the gluonic flux tube adapts to the quark and antiquark
positions. This is an adiabatic assumption,
in the quantum mechanical sense and not in the thermodynamic sense. 
Thus with the mechanical adiabatic assumption of a slow motion of the
flux tube, the quark movement is sufficiently slow in the heat bath of
the hot medium at temperature $T$, and we can assume that the
flux tube transformation is isothermal. Then the potential energy is well approximated by the free energy $V \simeq F$.
Nevertheless, possibly the quarks move too fast for a completely isothermal transformation of the flux tube, and then the flux tube transformation may have a small contribution from the internal energy $E$, identical to the
potential energy when no heat is exchanged with the heat bath,
in that case,
\be
V = (1- \omega) F + \omega E \ ,
\ee 
where $\omega$ should be a small number. 
In that case the potential dependence in the temperature $T$ is flatter
for $T<T_c$. In particular we also depict in Fig. \ref{flattening} the
cases where $ 0.1 < \omega <0.3$ where the string tension is
nearly flat up to close to $T=T_c$. 

Notice however that there is an evidence contrary to the use of the internal
energy $E$ for light quarks. The problem with the internal energy is not only it's divergence at $T=T_c$, but also that it is larger than the free energy, $E>F$. Yamamoto, Suganuma and Iida found in Lattice QCD that a the
presence of light quark does reduces the string tension
\cite{Yamamoto:2007pf}.
Thus, presently, the best model for the string tension at finite $T$ 
remains the one of Eqs. (\ref{eqformagnetization}) and (\ref{anssigma}).

\section{Conclusion}

We fit the finite temperature string tension  from the
free energies of a quark-antiquark pair computed in SU(3) lattice QCD by the
Bielefeld group
\cite{Doring:2007uh,Hubner:2007qh,Kaczmarek:2005ui,Kaczmarek:2005gi,Kaczmarek:2005zp}. 
We find a good ansatz for the string tension at all temperatures,
the magnetization curve of a ferromagnet. While
evidence for a 1st order phase transition exists
\cite{Kaczmarek:1999mm}, 
the difference to the 2nd order magnetization curve is small, 
and therefore the magnetization curve is a good ansatz for the 
quark-antiquark potential at finite temperature,
adequate to be included in the $\chi$QM.

We find that a constant string tension may also be arguably
used at $T<T_c$. Thus at  $T<T_c$ the quark model
computations present in the literature are acceptable. 
The main arguments are that the free energy $F$ string tension
is already relatively flat up to $T$ close to $T_C$ and that
the light hadron spectrum scales with $\sqrt \sigma$ which 
is flatter than the string tension, as in Fig. \ref{flattening}. 
Moreover, if a contribution
from the internal energy $E$ string tension can be justified,
then the string tension is further flattened  up to $T$ close 
to $T_C$. And the 1rst order discontinuity, although small,
also flattens the string tension.

However, at $T>T_C$ the string tension vanishes, and
in this sense the  $\chi$QM calculations present in the 
literature must be corrected. To encompass all temperatures,
the free energy string tension of Eqs. (\ref{eqformagnetization}) and (\ref{anssigma}) should be used in the $\chi$QM, 
or at least a step function with a transition at $T_c$ could be approximately used.

Nevertheless, the relative instability of the fits of the finite temperature
string tension, and the importance of the finite $T$ Coulomb or
logarithmic potentials suggest that eventually the finite $T$
 $\chi$QM will have to go beyond the simple use of
a long range linear potential, including also medium range 
and short range potentials
\cite{Bicudo:2008kc,Bicudo:2003cy}.

In what concerns the Lattice QCD studies of the free
energy at finite temperature, it would be interesting 
if more results would be further computed. In particular 
the subtle 1st order SU(3) discontinuity at $T=T_c$ 
\cite{Kaczmarek:1999mm},
deserves more detailed studies in the region of $T\simeq T_c$.
Moreover at smaller temperatures,  $T \in ]0, 0.8 T_c[$ there is
no Lattice QCD data. 
Thus we anticipate that the research of finite $T$ quark-antiquark
potentials will remain very interesting in the future.

 \hspace{2cm}
\acknowledgements
I thank Olaf Kaczmarek for sharing his SU(3) Lattice QCD data on the free energy, the internal energies and the free energy string tensions, 
Peter Petreczky for sharing his SU(2) Lattice QCD Data on the free energy string tension,  and Nuno Cardoso for his codes on linear and non-linear fits. 
I am grateful to Marlene Nahrgang, to Olaf Kaczmarek, to Pedro Sacramento, to Jan Pawlowski and to Fabien Buisseret for discussions on the QCD phase diagram motivating this paper.  I acknowledge the financial support of the FCT grants CFTP, CERN/FP/109327/2009 and CERN/FP/109307/2009.


\end{document}